\documentclass[aps,prl,twocolumn,groupedaddress,superscriptaddress,showpacs]{revtex4-1}
\usepackage{epsfig}

\begin{document}

\title{Impact of Perturbations on Watersheds}

\author{E. Fehr}
\email{ericfehr@ethz.ch}
\author{D. Kadau}
\affiliation{IfB, ETH Z\"urich, 8093 Z\"urich, Switzerland}

\author{J. S. Andrade Jr.}
\author{H. J. Herrmann}
\affiliation{IfB, ETH Z\"urich, 8093 Z\"urich, Switzerland}
\affiliation{Departamento de F\'{\i}sica, Universidade Federal do Cear\'a, 60451-970 Fortaleza, Cear\'a, Brazil}

\begin{abstract} 
  We find that watersheds in real and artificial landscapes can be
  strongly affected by small, local perturbations like landslides or
  tectonic motions. We observe power-law scaling behavior for both the
  distribution 
  of areas enclosed by the original and the
  displaced watershed as well as the probability density to induce,
  after perturbation, a change at a given distance. Scaling
    exponents for real and artificial landscapes are determined, where
    in the latter case the exponents depend linearly on the Hurst
    exponent of the applied fractional Brownian noise. The obtained
    power-laws are shown to be independent on the strength of
    perturbation. Theoretical arguments relate our scaling laws for
    uncorrelated landscapes to properties of invasion percolation.
\end{abstract}


\maketitle 

Watersheds are the lines separating adjacent drainage basins
(catchments) and play, hence, a fundamental role in water management
\cite{Vorosmarty98}, landslides \cite{Dhakal04,Lee06}
and flood prevention \cite{Lee06,Burlando94}.
Since ancient times watersheds have been used to delimit boundaries and
have already become issues in disputes between countries
\cite{UN1902}. Moreover, similar problems also appear in other areas
such as Image Processing and Medicine \cite{Vincent91},
which shows the generality and importance to fully understand the
subtle dynamical properties of watersheds. But how sensitive are
watersheds to slight localized modifications of the landscapes? Can
these perturbations produce large, non-local changes in the watershed?
Geographers and geomorphologists have studied the evolution of
watersheds in time and found it to be driven by local events called
stream capture. These events can affect the biogeography \cite{Burridge07},
and occur due to erosion, natural damming, tectonic motion as well as volcanic activity
\cite{Garcia-Castellanos09,Linkeviciene09,Dorsey06,Bishop95}. Recently, the associated relevant
mechanisms were investigated numerically and in small scale experiments \cite{Attal08}.
Finally, the problem studied here is also of interest to image processing, in order
to circumvent segmentation failure \cite{Patil09}.

\begin{figure}[!ht] 
\centerline{\epsfig{figure=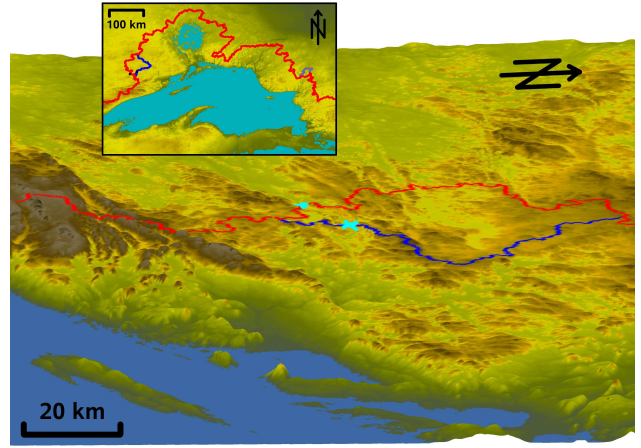,width=8.3cm}}
\caption{(color online) Example of the watershed between US and Canada, close to the
  big lakes (red). Also shown is the resulting change in the watershed
  (blue) due to a perturbation of $2$ m at a spot (cross) close to the
  border, near Thunder Bay. The watershed displacement encloses an
  area of about $3730$ km$^2$.  The dot marks the new \emph{outlet} of
  the area after perturbation. The inset shows the same area on a
  larger scale.}\label{fig1}\vspace{-3ex}
\end{figure}

In this Letter we investigate the effects of topological
modifications like landslides or tectonic motion on the
watershed. In fact, we show that the same type of topological
perturbation can indeed trigger non-local effects of any length
scale, i.e., following power-laws distributions. For illustration,
as shown in Fig.~\ref{fig1}, we obtain after a local height change
of less than 2~m at a location (cross) close to the Kashabowie
Provincial Park, some kilometers North of the US-Canadian border, a
substantial displacement in the watershed (blue), which
encloses together with the original watershed (red) an area $A\sim
3730$ km$^2$. Here a model is developed to provide a qualitative
and quantitative description of this phenomenon.

In our simulations, we use real and artificial landscapes in the form
of Digital Elevation Maps (DEM), consisting of discretized elevation
fields. Here we call \emph{sites} to the discretization units,
defined as the square areas with a size given by the DEM resolution.
The watershed is the line dividing the entire landscape into two
parts. Each part drains, according to the steepest descent along the
coordinate directions, to either one of a chosen pair of opposite
boundaries (east-west or north-south) of the DEM. For the
determination of this line we use an iterative application of an
invasion percolation procedure (IP) \cite{Fehr09}. For a given
landscape, as shown in Fig.~\ref{fig1}, we initially determine its
watershed (red line). Then a local event is induced by changing the
height $h_k\rightarrow h_k+\Delta$ at a single site $k$ (cross in
Fig.~\ref{fig1}) of the DEM, where $\Delta$ is the perturbation
strength. Since we are interested on the non-local features of
the watershed response to local perturbations, we only perturb sites
that are not on the watershed. This implies $\Delta>0$ to induce
changes in the watershed (blue line). The displacement of the
watershed is quantified by measuring the area $A$ between the original
and the perturbed watershed. By definition, the water can only escape
from the displacement area through a single site, which we call
\emph{outlet}. The old outlet ($o$-outlet) before perturbation
always coincides with the perturbed site $k$. After perturbation,
$k$ becomes part of the new watershed and the water escapes through
a new outlet ($n$-outlet), which is located at the original
watershed. After measuring the area $A$ and the distance $R$ between
the old and new outlets, we proceed by restoring the original
landscape, i.e., the height at $k$ is reset to its initial value.
This procedure is repeated for every site $k$ of the landscape,
except those located at the original watershed. Initially, we
fix $\Delta$ equal to $h_w=|h_{max}-h_{min}|$, i.e., the height
difference between the lowest $h_{min}$ and highest height $h_{max}$
of the landscape, which corresponds to a perturbation of infinite
strength for the landscape under investigation. With this choice all
possible changes are obtained within the DEM. In all definitions
hereafter, we consider only those perturbations leading to a
displacement of the watershed. In what follows, we study the
distribution 
$P(A)$ of the areas $A$, the probability
distribution $P(R)$ of the Euclidean distance $R$ between the two
outlets, and the dependence between $A$ and $R$.  For this, we define
the average area $\left< A\right>$ and the distribution 
$P(A|R)$ of areas $A$ associated with an outlet distance $R$.

\begin{figure}[!ht] 
\centerline{\epsfig{figure=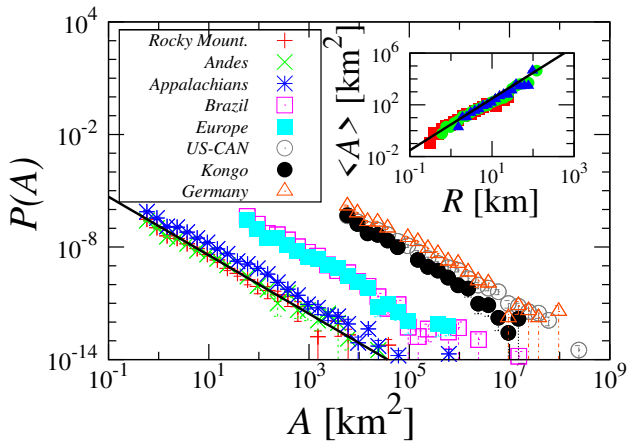,width=8.3cm}}
\caption{(color online) The distribution 
$P(A)$ is shown for various regions:
Rocky Mountains, Andes and Appalachian (unshifted); Brazil and
Europe (shifted by a factor of $10^2$ for better visibility); US-CAN,
Kongo and Germany (shifted by $10^4$). All data sets have a
resolution of $540$~m. The solid line shows the fit to the
Andes data with a power-law of exponent $-1.65 \pm 0.15$. The inset
shows $\left< A\right>$ as function of $R$ for the Rocky Mountains at
resolutions of $270$~m (squares), $540$~m (circles) and $1350$~m
(triangles). The solid line has slope 2.}\label{Naturalfig}
\vspace{-2ex}
\end{figure}

First, we study several natural landscapes, from mountainous
(e.g. Rocky Mountains) to rather flat landscapes (e.g. US-CAN, Kongo
and Germany). The DEM data was obtained from the SRTM-project
\cite{Farr07}, where for each set we used a size of $2700$
km$\times 2700$ km (except $1080$ km$\times 1080$ km for Germany),
and a resolution of $540$~m, defining the size of a site.
Hence, the physical size of the 8 data sets are large enough, so
that finite size effects emerging from the DEM boundaries could not be detected.
As shown in Fig.~\ref{Naturalfig}, we find the
distribution 
of areas to follow a power-law, $P(A)\sim
A^{-\beta}$, with $\beta= 1.65\pm 0.15$ for all
landscapes. The probability distribution $P(R)$ of outlet distances
$R$ also obeys a power-law, $P(R)\sim R^{-\rho}$, with $\rho= 3.1\pm
0.3\approx 2\beta$ (see Fig.~SM1 in \cite{supplementary}), and displays an upper cutoff in the range $50$ km
$< R < 500$ km for the studied landscapes. This cutoff is
independent on the resolution and could be due to a length scale
arising from tectonics. The value of $\rho$ implies $\left< A\right>\sim
R^{2}$, which agrees well with our data (inset of
Fig.~\ref{Naturalfig}). The distribution 
for a given distance
$R$ scales as $P(A|R)\sim A^{-\alpha}$ with $\alpha = 2.3\pm 0.2$ (see Fig.~SM1 in \cite{supplementary}).

\begin{figure}[!ht] 
\centerline{\epsfig{figure=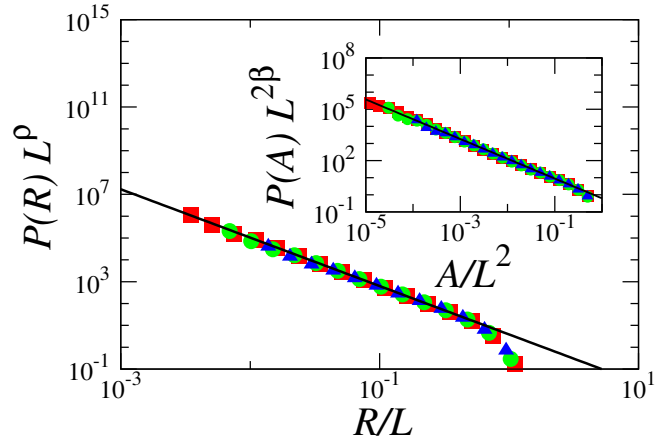,width=8.5cm}}
\caption{(color online) Data collapse of the distribution $P(R)$ for
uncorrelated landscapes ($H=-1$) of three different system sizes
$L=129,\, 257,\, 513$ (triangles, circles and squares,
respectively). The line represents a power-law fit to
the data for the largest landscape (squares) revealing an exponent
$\rho=2.21\pm 0.01$. The inset shows the distribution 
$P(A)$ of
the areas for the same system sizes. The line represents a power-law
fit to the data with exponent $\beta=1.16\pm 0.03$.}\label{Uncorrfig1} \vspace{-2ex}
\end{figure}

In order to understand these power-laws and the dependence
between $A$ and $R$, we study artificial landscapes, where the local
heights are generated using fractional Brownian motion (fBm) on a
square lattice \cite{Lauritsen93}. This model incorporates spatial long-range
correlations to the system that are controlled by the Hurst
exponent, $H$. We first consider the case of uncorrelated
landscapes, which is a special case of the fBm model with
to $H=-1$. In Fig.~\ref{Uncorrfig1}, we present the
results obtained for several system sizes, using the same procedure as
for the natural landscapes.  The probability density $P(R)$
again follows a power-law $P(R)\propto R^{-\rho}$, without
upper cutoff as in real landscapes. We estimate $\rho=2.21\pm
0.01$ using the scaling $P(R)=L^{\rho} f[RL]$, where $L$ is the
linear dimension of the landscape. For the distribution 
$P(A)=L^{\beta} f[AL^{2}]$, we obtain an excellent data collapse for
$\beta=1.16\pm0.03$ (see the inset of Fig.~\ref{Uncorrfig1}).  In the
case of the distribution $P(A|R)$ at a fixed outlet distance, we again
find a power-law $P(A|R)\sim A^{-\alpha}$ (see Fig.~SM2 in \cite{supplementary}). Finite
size scaling analysis yields an exponent $\alpha = 2.23\pm 0.03$
independent on the value of $R$. Assuming that $R$ describes the
extension of $A$ in every direction, the relation $\rho=\alpha$ is
reasonable. This is even well supported by the similarity of the
obtained exponents. The area $A$ was rescaled by $L^2$, indicating
that the areas are compact. Considering finite-size scaling, our data is consistent
with $\left< A\right>\sim R^2$ (see Fig.~SM2 in \cite{supplementary}).
Furthermore, the compactness of the areas is
supported by the measured value $\beta=1.16\pm 0.03$, which agrees well
with the relation $\beta=\rho/2\approx 1.11$.

\begin{figure}[!ht] 
\centerline{\epsfig{figure=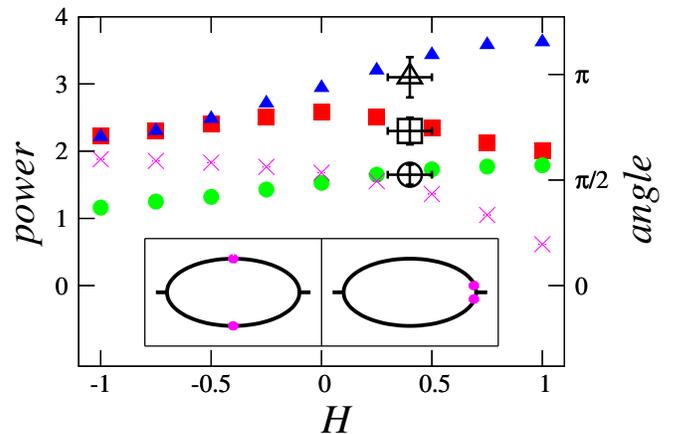,width=8.5cm}}
\caption{(color online) The exponents $\alpha$ (squares), $\beta$ (circles) and
$\rho$ (triangles) are shown for several values of the Hurst
exponent $H$.  Each point results from a similar study as done for
the uncorrelated landscapes. The exponents for the natural
landscapes (open symbols), all corresponding to Hurst exponent
values in the range $0.3<H<0.5$, are consistent with our model.
The average angle $\theta$ (in radians) between the outlets from the
center of mass is shown too (crosses). The insets depict schematic
shapes of the areas and positions of the two outlets for small
(left) and large (right) values of $H$.}\label{fBmfig} \vspace{-2ex}
\end{figure}

In the following we show that we can match the exponents
quantitatively by tuning the Hurst exponent $H$ to introduce spatial
long-range correlations, as present in real geological systems.
The exponents $\alpha$, $\beta$ and $\rho$ were calculated for several
values of $H$ (see Fig.~\ref{fBmfig}). As shown in
Fig.~\ref{fBmfig}, we observe that both $\beta$ and $\rho$ increase
with $H$. Furthermore, the relationship $\beta=\rho/2$ is
maintained, since the areas remain compact in the entire range of
$H$ values.  Around $H=-0.5$, $\alpha$ starts to deviate from $\rho$
and for $H>0$ we observe $\alpha$ to decrease. Previously, we had
assumed $R$ to reflect the extension of the area, i.e., the outlets
to reside typically on opposite sides of the area. To check whether
this is still valid, we measured the angle $\theta$ between the
lines connecting the center of mass of the area with the two
outlets. We observe the average angle to decrease as a function of
$H$ (see Fig.~\ref{fBmfig}). This implies that, on average, the two outlets
approach each other with increasing $H$ (see also the insets of
Fig.~\ref{fBmfig}), so that $R$ is no longer representative of the
area extension. Finally we find good quantitative agreement
with the exponents obtained from the natural landscapes, which are
known to have a Hurst exponent inside the range $0.3<H<0.5$ (see
Ref.  \cite{Pastor-Satorras98} and references therein). Hence,
except for the upper cutoff in $R$, our model provides an
excellent quantitative description of the effects observed on natural
landscapes.
\begin{figure}[ht] 
\centerline{\epsfig{figure=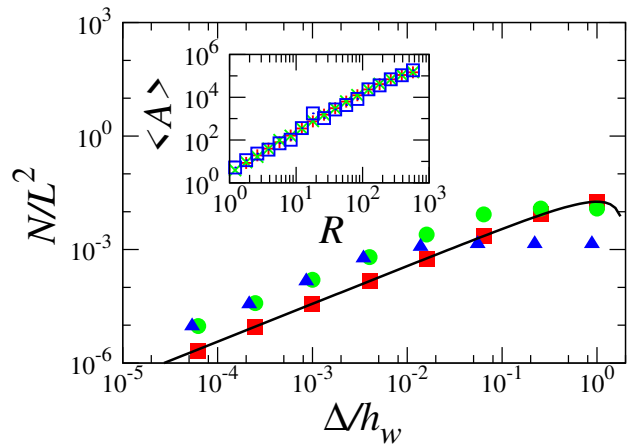,width=8.5cm}}
\caption{(color online) Dependence of the number of perturbed sites $N$ that promote
changes on the watershed on the perturbation strength $\Delta$
applied for uncorrelated (squares), Andes (triangles) and fBm
landscape with $H=0.3$ (circles). The solid line corresponds to the
analytic relation obtained from Eq.~(\ref{Nps}) for uncorrelated
landscapes. The inset shows the average area $\left< A\right>$ as a function
of the distance $R$ between the outlets for $\Delta/h_w=1,\, 0.016$
and $0.00025$ (pluses, crosses, and squares, respectively), and
$L=513$.}\label{NDeltafig}
\vspace{-2ex}
\end{figure}

Next we analyze quantitatively the impact of the perturbation strength
$\Delta$ on the watershed. In Fig.~\ref{NDeltafig} the number of
perturbed sites $N$ that change the watershed is shown for
uncorrelated, for artificial correlated (with $H=0.3$) and for natural landscapes
(Andes). In all three cases, $N$ is found to increase linearly with
the applied perturbation strength, $N\sim \Delta$. This indicates that
changes on the watershed can be observed even for infinitesimally
small perturbations. Additionally, in both cases where correlations
are present, $N$ is observed to reach a plateau. As already stated,
when $\Delta$ is equal to $h_w$, this corresponds to the largest
relevant perturbation, so that $N(h_w)=N_{max} \ll L^2$ indicates
that many perturbations never change the watershed at all. It is
clear that $\Delta>|h_j-h_i|$ is needed, where $h_i$ and $h_j$ are
the heights of the outlets of the area. Therefore, if the
distribution $p_o(h)$ of outlet heights is known, one obtains,
\begin{equation}\label{Nps}
  N(\Delta) = 2\frac{N_{max}}{L^2}\, \int_0^{\Delta} \int_{\Delta'}^{h_w}p_o(h)\, 
  p_o(h-\Delta')\, \mathrm{d}h\,\mathrm{d}\Delta'\,\, .
\end{equation}
For landscapes with uniformly distributed heights, we find $p_o(h)$ to
be still a uniform distribution. Then we obtain from Eq.~(\ref{Nps}),
$N(\Delta)=(h_w\Delta-\Delta^2/2) 2N_{max}/(L^2h_w^2)$, which is in
excellent agreement with our data (see Fig.~\ref{NDeltafig}), where an
approximately linear behavior can be observed for $\Delta < h_w$. The
observed power-laws are maintained for all values of $\Delta$, as can be 
clearly seen for $\left< A\right>$ in the inset of Fig.~\ref{NDeltafig}.
We conclude that infinitesimally small
perturbations have qualitatively the same effect on the watershed as
any larger perturbation strength $\Delta$.

In the case of uncorrelated landscapes ($H=-1$), for a given area $A$,
the corresponding invasion percolation cluster is obtained by starting
the penetration process from one outlet to another, always growing
along the steepest descent. The area $A$ can therefore be understood
as the envelop of this IP-cluster. From percolation theory, the
fractal dimension of the IP cluster is $d_{f}=91/48$ in two dimensions
\cite{Araujo05}, which implies that $\left<A\right> \propto
M^{2/d_{f}}$, where $M$ is the number of sites (mass) of the cluster.
This result is consistent with our simulations. The size-distribution
$P(M|R)$ of IP-clusters between two sites at a fixed distance $R$ is
known to follow a power-law $M^{-\alpha^{*}}$ with $\alpha^{*}=1.39$
\cite{Araujo05}. Note that, for comparison of our results to Ara\'ujo
{\it et al.} \cite{Araujo05}, $P(M|R)$ needs to be divided by $M$, as
we grow the IP-cluster starting from the outlet at the watershed,
which is always the highest of the $M$ sites of the cluster. Hence we
expect $P(M|R)\sim M^{-(\alpha^{*}+1)}$, what is indeed in good
agreement with our data (see Fig.~SM3 in \cite{supplementary}).
We can now relate our exponent $\alpha$ of the distribution of areas at fixed distance to
$\alpha^{*}$, which describes the size-distribution
of IP-clusters, as
$P(A|R)=P(\left< A\right>(M)|R)\propto \left< A\right>^{-\alpha}(M)\propto M^{2 \alpha
 /d_f}\propto P(M|R)$. We obtain
$\alpha=\frac{d_f}{2}(\alpha^{*}+1)\approx 2.266$, which is very close
to what we measure ($\alpha=2.23\pm 0.03$). Therefore, we can relate
our results on uncorrelated landscapes to the sub-critical point-to-point
invasion percolation process \cite{Araujo05} and to the mass
distribution of avalanches that occur during the IP-cluster growth
\cite{LomnitzAdler92,Amaral95,Ferer02}.

In summary, we were able to show that small and localized
perturbations can have a large impact on the shape of watersheds even
at very long distances, hence having a non-local effect. The
distribution 
of changes $P(A)$ is found to decrease as a
power-law with exponent $\beta=1.65\pm 0.15$ on all studied real
landscapes from mountainous (e.g. Rocky Mountains) to rather flat
(e.g. US-Canadian border). By applying perturbations to model landscapes
with long-range correlations, we determined the dependence of the
scaling exponents on the Hurst exponent, finding good quantitative
agreement with real landscapes, for which $0.3<H<0.5$. The
obtained exponents $\alpha,\,\beta$ and $\rho$ are independent of the
perturbation strength $\Delta$. For uncorrelated landscapes, we derived
a relation with invasion percolation. It is known that watersheds
\cite{Fehr09} on uncorrelated landscapes are related to ``strands'' in
Invasion Percolation \cite{Cieplak96}, random polymers in strongly
disordered media \cite{Porto99}, paths on MST's \cite{Dobrin01}, the
backbone of the optimal path crack \cite{Andrade09} and the cluster
perimeter in explosive percolation \cite{Araujo10}. Hence, our results
can be potentially applied to all these problems.

We acknowledge useful discussions with N. A. M. de Ara{\'u}jo and
thank CNPq, CAPES, FUNCAP, and the
CNPq/FUNCAP-Pronex grant for financial support.

\end{document}